\documentclass[aps,prl,twocolumn,superscriptaddress,showpacs,]{revtex4-1}

\usepackage{amsmath}
\usepackage{amssymb}
\usepackage{graphics}
\usepackage{graphicx}
\usepackage{epstopdf}
\usepackage{dcolumn}
\usepackage{bm}
\usepackage{color}

\newcommand{\beq}{\begin{equation}}
\newcommand{\eeq}{\end{equation}}
\newcommand{\mtE}{\mathcal{E}}
\newcommand{\lgl}{{\langle}}
\newcommand{\rgl}{{\rangle}}

\begin{document}

\title{Rogue Waves, Self-similarity, and Integrable Turbulence}

\author{Yashar E. Monfared and Sergey A. Ponomarenko}
\affiliation{Department of Electrical and Computer Engineering, Dalhousie University, Halifax, Nova Scotia, B3J 2X4, Canada}
\date{\today}

\begin{abstract}
We explore extreme event occurrence in the integrable turbulence with self-similar asymptotics. We posit that rogue waves in such systems  manifest themselves as giant fluctuations away from average self-similar dynamics of the system. We support our proposition with numerical simulations of rogue wave excitation in the self-similar regime of stimulated Raman scattering. We show that our results hold irrespective of a specific source correlation model, suggesting the universality of the proposed scenario.

\begin{description}

\item[PACS numbers] {42.65.-k, 42.65.Sf, 42.65.Dr}

\end{description}

\end{abstract}

\maketitle

\noindent{\emph{Introduction}.--} Turbulence, defined as chaotic changes of dynamical variables throughout the evolution of the corresponding physical systems, is a subject with a long and venerable history~\cite{Batch}. The subject has recently acquired a new dimension with the introduction of the integrable turbulence concept by Zakharov~\cite{Zakh} who pointed out that nonlinear statistical wave systems, described by the integrable equations, differ from conventional turbulent systems in several important aspects. In this context, the rogue wave (RW), or, in general, extreme event excitation mechanisms in integrable turbulence present a special interest. RWs are extremely rare, giant waves, obeying non-Gaussian statistics~\cite{RW-rev}. To date, RWs have been proven ubiquitous in oceanography~\cite{Dya}, plasma physics and Bose-Einstein condensates~\cite{RW-rev} as well as in nonlinear optics~\cite{Solli,Dud-rev} among many other branches of physics.

The RW generation in integrable turbulence is commonly studied within the framework of a generic $(1+1)$D nonlinear Schr\"{o}dinger equation (NLSE). The latter governs time-evolution of weakly dispersive wave systems with weak instantaneous nonlinearities of the Kerr type~\cite{Agra}. The RW excitation in the NLSE model with random input wave fields has been studied both numerically~\cite{Pic,Dud,Aga,Akh1, Akh2,Sur1,Sur2} and experimentally~\cite{Sur1,Sur2}. These studies helped elucidate the respective roles of  spontaneous Peregrine-like breather excitation from a noisy environment and of random soliton collisions in triggering the emergence of heavy-tailed probability density distributions (PDF) of field intensities. Such heavy-tailed PDFs herald the RW generation in the system~\cite{Dud,Sur1,Sur2,Akh1,Akh2}. 

However, the NLSE model fails to accurately describe the dynamics of nonlinear waves in the vicinity of either wave-wave or wave-matter resonances.  The resonant wave-wave or wave-matter interaction regime is characterized by strong nonlinearity, long coherent memory of the system, and strong amplification/absorption near resonance~\cite{Ebly}. In optical physics, the resonant RW  excitation can be generically modelled through either two-level amplification (TLA) or stimulated Raman scattering (SRS) processes. The energy transfer to an optical wave from either (inverted) medium atoms (TLA) or from another wave via two-photon resonance (SRS) serves as a generic resonant amplification mechanism. In fluid mechanics, a resonant interaction between short surface and long internal waves occurs whenever the phase velocity of a short wave matches the group velocity of a long wave.
The short-wave-long-wave interaction equations are known to be integrable~\cite{Jap}. The TLA and SRS are also governed by integrable equations in the transient regime~\cite{Lamb,Chu,Kaup}. The key feature of both TLA and SRS is a transient character of solitons and breathers and universal long-term self-similar evolution of the system in the integrable limit~\cite{Myk}. Thus, neither soliton collisions nor breathers can trigger extreme events in the long-term evolution of self-similar integrable turbulence. A fundamental question then arises: What is a physical manifestation of extreme events and, especially, RWs if any, in  the {\it integrable turbulence with self-similar asymptotics?}

In this work, we explore extreme event excitation in the long-term, self-similar regime of integrable turbulence for the first time to our knowledge. We propose that RWs appear as rare, giant fluctuations away from a self-similar, on average, dynamics of the system. Our extensive numerical simulations of extreme event occurrence in the self-similar regime of SRS reveal the existence of RWs and support the proposed scenario of their emergence. We also confirm that RWs can be excited regardless of the  source statistical model. We anticipate the proposed RW excitation scenario to be ubiquitous for self-similar integrable turbulence, characterized by long coherent memory.

\noindent{\emph{Self-similar regime of SRS}.--}We consider the Raman interaction between a pump and fundamental Stokes modes. This SRS modality can be realized, for instance, in a gas-filled hollow core photonic crystal fiber, designed to suppress higher-order Stokes modes~\cite{Naz}. In the extreme transient regime, the characteristic SRS interaction time $T_{\mathrm{SRS}}$ is much shorter than the medium dipole relaxation time $T_2$~\cite{Yash1}. The slowly-varying pump $\mtE_p$ and Stokes $\mtE_s$ pulse amplitude evolution is then governed by simplified Maxwell's equations
	\beq\label{Max}
		\partial_{Z}\mtE_p = -\kappa\sigma\mtE_s, \hspace{1cm} 
			\partial_{Z}\mtE_s = \kappa^{-1}\sigma\mtE_p,
		\eeq
and the dipole moment matrix element $\sigma$ obeys the Schr\"{o}dinger equation, which, in the weak excitation approximation pertaining to realistic experimental conditions, takes the form~\cite{Yash1,Yash2,Chu,Kaup,Myk}
	\beq\label{Sch}
		\partial_{T}\sigma=\mtE_p \mtE_{s}.
		\eeq
Here $\kappa = \sqrt{\omega_p n_s / \omega_s n_p}\approx 1$, $\omega_{p,s}$ and $n_{p,s}$ being the carrier frequencies and linear refractive indices of the pump and Stokes modes, respectively. In writing Eqs.~(\ref{Max}) and~(\ref{Sch}), we assumed chirpless pump and Stokes input pulses, implying that all field variables are real, and used the same dimensionless variables as in Ref.~\cite{Yash1,Yash2}.  
\begin{figure}[t]
\centering
\includegraphics[width=\columnwidth]{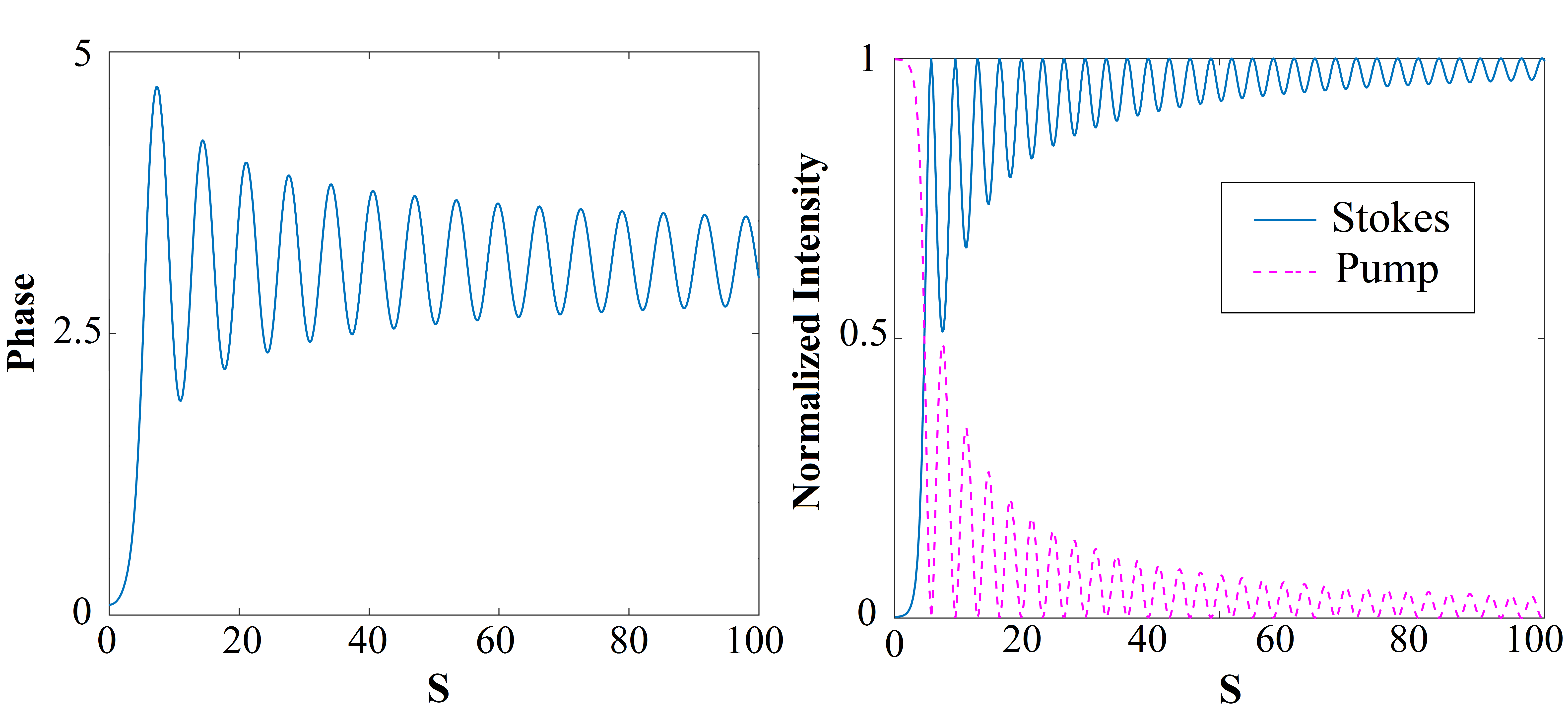}
\caption{(color online). (Left panel:) Self-similar phase $\theta$ as a function of the similarity variable $s$. We take $\kappa=1$ and assume that the energy initially resides almost entirely with the source pump mode, $P_{0s}=0.01 P_{0p}$. (Right panel:) Normalized intensities of the Stokes (solid blue) and pump (dashed magenta) pulses as functions of $s$. }
\label{figure1}
\end{figure}

The power conservation law implies the following parametrization of the pulse amplitudes
	\begin{subequations}\label{SRS-pst}
	\beq
		\mtE_p =(\kappa K)^{1/2}\cos(\theta/2),
		\eeq
and
	\beq
		\mtE_s =(K/\kappa)^{1/2}\sin(\theta/2).
			\eeq
			\end{subequations}
Here $\theta$ is a real phase and the integral of motion $K(T)$ is defined as
	\beq\label{Pow}
		K(T)= \kappa^{-1}\mtE_p^2(T,0) +\kappa\mtE_s^2 (T,0).
			\eeq
In SRS in molecular gases, $\kappa\approx 1$ and $K(T)$ is well approximated by the total intensity at the source, $I_{\mathrm{tot}}(T,0)$. In the self-similar regime, the phase depends only on the similarity variable $s$ such that~\cite{Elg,Myk1}
	\beq\label{SS}
		\theta=\theta(s), \hspace{1cm} s=2\sqrt{Z\int_{-\infty}^{T}dx\,K(x)}.
			\eeq
The SRS dynamics is then governed by the single ordinary differential equation in the form
	\beq\label{SRS-SS}
		\theta_{ss}^{\prime\prime}+\textstyle\frac{1}{s}\theta_{s}^{\prime}=\sin\theta.
			\eeq
\begin{figure}[t]
\centering
\includegraphics[width=\columnwidth]{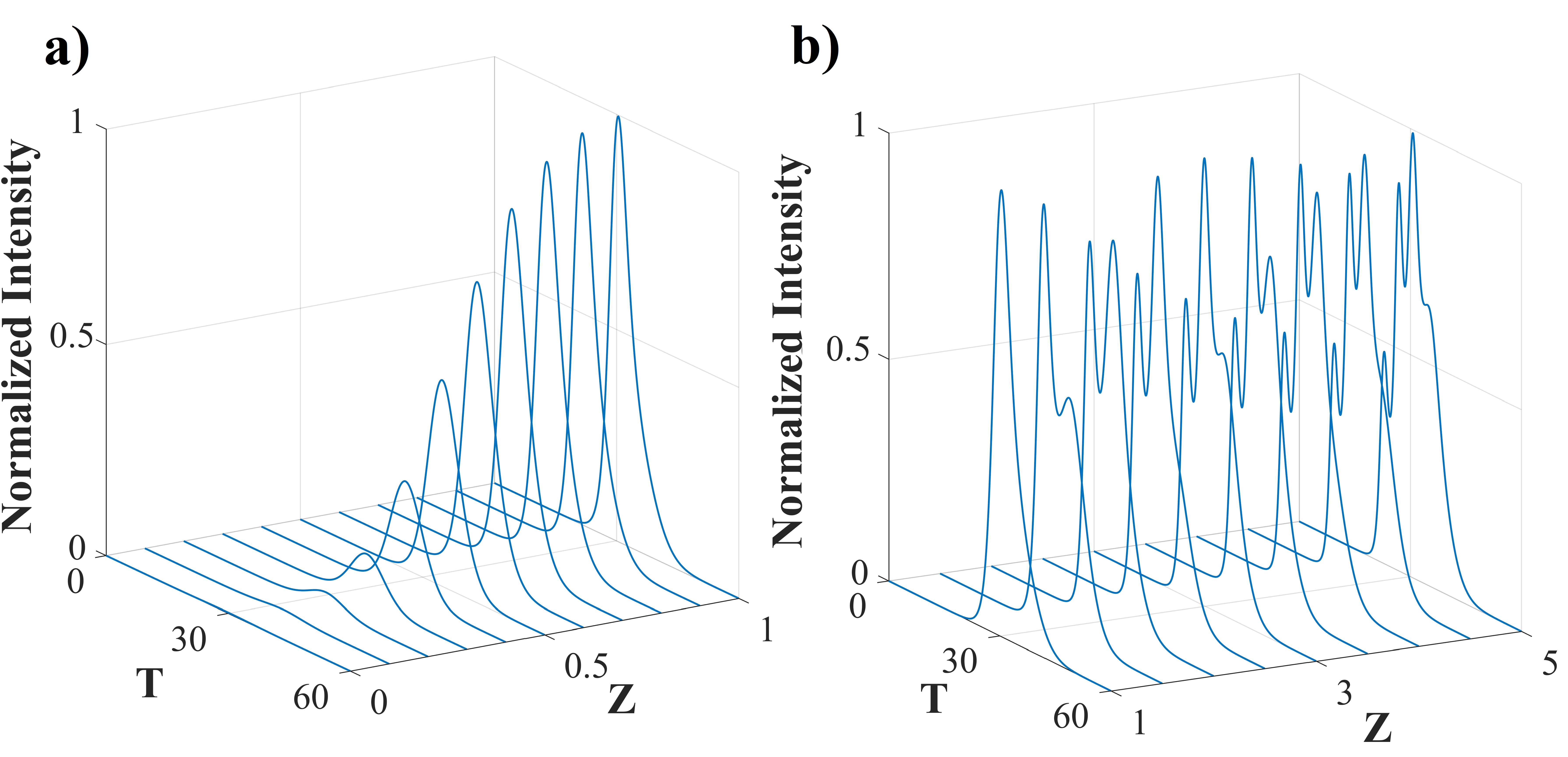}
\caption{(color online). Evolution of the normalized intensity of a Gaussian Stokes pulse with no input noise at the source as a function of dimensionless time $T$ and propagation distance $Z$ for two stages: (a) initial exponential amplification stage and (b) self-similar stage. The other parameters are: $\kappa=1$, $T_0=30$ and the Gaussian pump/Stokes pulse duration is $T_{\ast}=8.3$.}
\label{figure2}
\end{figure}
Eq.~(\ref{SRS-SS}) must be solved subject to the boundary conditions, $\theta^{\prime}(0)=0$ and $\theta(0)=\tan^{-1}(\kappa\sqrt{n_p P_{0s}/n_s P_{0p}})$, where $P_{0s}$ ($P_{0p}$) is an input power of the Stokes (pump) pulse.  Hereafter, we assume that the energy initially resides almost entirely with the source pump mode, $P_{0s}=0.01 P_{0p}$. In Fig. 1,  we exhibit a numerical solution to Eq.~(\ref{SRS-SS})  (left panel) and the universal normalized intensity profiles of the pump, $\overline{I}_p=I_p/\kappa K$ (magenta curve, right panel) and Stokes,
$\overline{I}_s=\kappa I_s/K$ (blue curve, right panel).  We clearly observe gradual pump mode energy depletion, resulting in the Stokes mode amplification in Fig. 1. The process is accompanied by coherent oscillations due to long coherent memory of the SRS interaction in this regime. In Fig.~2, we display the normalized intensity evolution of a Stokes pulse with no input noise at the source. The input Stokes and pump pulses are assumed to have the same Gaussian profile with the temporal width $T_{\ast}=8.3$, which corresponds to a physical pulse duration of $t_{\ast}=10$ ns and the SRS interaction time $T_{\mathrm{SRS}}=1.2$ ns~\cite{Yash2,Naz}; the pulses are centred at $T_0=30$.  The analysis of Fig. 2 reveals that an initial, nearly exponential amplification stage (Fig. 2a)  is followed by a self-similar evolution stage (Fig. 2b). The self-similar stage onset is unequivocally marked by the appearance of intensity oscillations. We then infer from the figure that the self-similar regime starts past around $Z_{cr}\simeq 1$ in our units. To ensure that our system is well into the self-similar regime, we examine extreme event excitation in the Stokes output at distances larger than $Z_{cr}$.

\begin{figure}[t]
\centering
\includegraphics[width=\columnwidth]{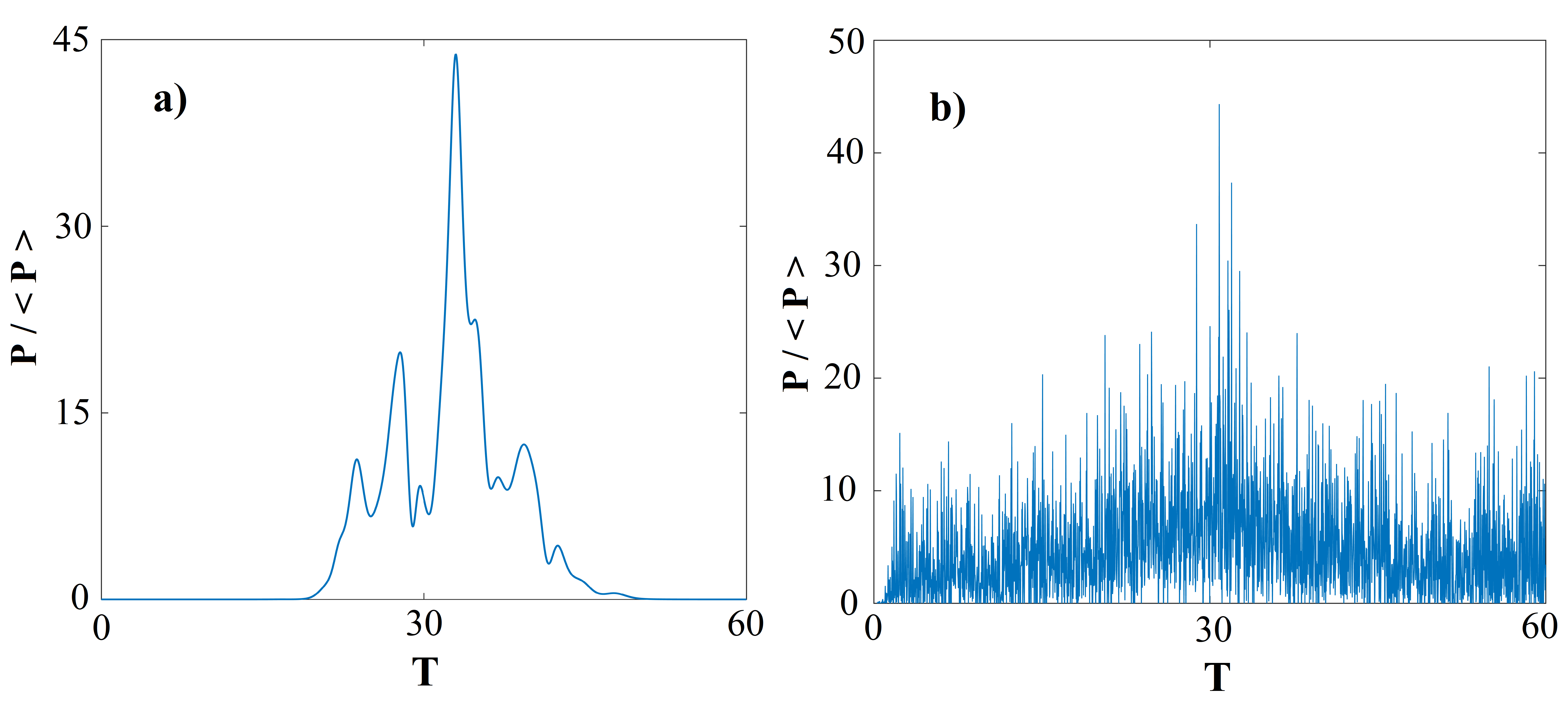}
\caption{(color online). Power fluctuations of a random realization of the Stokes pulse ensemble at $Z=3$ for (a) GSM and (b) FBWN source pump pulse ensembles. The numerical parameters are: $\Delta P/P_0=0.4$, $\kappa=1$, and $T_c=0.1T_{\ast}$. The Stokes pulse power is normalized to its average value at the same propagation distance, $Z=3$. }
\label{figure3}
\end{figure}

\noindent{\emph{Extreme events in self-similar integrable turbulence}.--}We study SRS with a small-amplitude coherent Stokes pulse seed, interacting with a strong fluctuating pump pulse. The pump pulse field at the source is assumed to consist of a coherent and random components such that
	\beq
		\mtE_p (T,0)=\mtE_{p0}(T,0) +\Delta\mtE_p(T,0),
			\eeq
where the random component can be expressed in terms of coherent modes $\{\psi_n(T)\}$ of the source as
	\beq
		\Delta\mtE_{p}(T,0)=2^{-1/2}\left(\sum_{n}c_{n}(T)\psi_{n}(T) + c.~c.\right).
		\eeq
Here $\{c_n\}$ are random complex amplitudes and $c.~c.$ stands for a complex conjugate. 

To explore the generality of our results, we consider two physically very different statistical models of the pump. In the first model, the coherent  component has a Gaussian pulse envelope centred at $T_0$ with the temporal duration $T_{\ast}$ (in the dimensionless units).  The random component has a Gaussian average intensity profile of the same duration, for simplicity, and a Gaussian fluctuation spectrum of the width inversely proportional to the source coherence time $T_c$. Thus, the pair correlations among the monochromatic components of such a pump source strongly decay with the frequency separation between the components. This model is known as a Gaussian Schell-model (GSM) which has been extensively employed before in statistical nonlinear optics~\cite{Yash1,Yash2,Mokh}. The second model allows for an arbitrary temporal envelope of the pump pulse, but assumes, for simplicity, that the random component has the same average intensity profile as the coherent component profile. Further, the model stipulates that the fluctuation spectrum of the random component be flat within a finite bandwidth. In this model, the spectral correlations among pairs of monochromatic components of the source are uniform within the source bandwidth, making this finite-bandwidth white noise (FBWN) model drastically different from GSM. In the Supplemental Material~\cite{Suppl}, we present details of statistical ensemble construction for both models. In particular, we are able to analytically determine the coherent modes $\{\psi_n(T)\}$ and demonstrate how the statistical properties of $\{c_n\}$'s shall be specified to ensure the source GSM and FBWN ensembles obey Gaussian statistics for any second-order coherence time. 

In the self-similar regime of interest, the Stokes and pump field ensemble representations are determined from Eqs.~(\ref{SRS-pst})  with their evolution being governed by Eq.~(\ref{SRS-SS}). We assume that the coherent pump pulse component for both GSM and FBWN ensembles has a Gaussian pulse profile and perform extensive Monte-Carlo simulations for an ensemble of $10^4$ realizations of the pump field. 
Hereafter, we assume that $\kappa =1$. We introduce an average peak power $\Delta P$  of the random component of the pump pulse at the source and the corresponding peak power of the coherent component $P_0$~\cite{Suppl}; their ratio is taken to be $\Delta P/P_0=0.4$ henceforth. In Fig.~3, we exhibit the normalized power of a Stokes pulse ensemble realization at a distance $Z=3$, well within the self-similar evolution domain for both GSM (Fig. 3a) and FBWN (Fig.3b) source ensembles; the Stokes pulse power is normalized to its average magnitude at $Z=3$. We notice a quantitatively different behaviour of the output Stokes pulse fluctuations in the two cases: the FBWN ensemble fluctuations are more spread in time than are the GSM ensemble ones. This is because the Gaussian GSM source time correlations are more localized than are the FBWN ones, the latter being governed  by a sinc function, (see the Supplemental Material for details.)  Further,  we can infer from the figure that the output Stokes pulse realization for either ensemble attains a peak power nearly 45 times its average power at this propagation distance, thereby unambiguously qualifying the output as an RW. Thus, RWs can be generated in the self-similar regime of SRS which is in qualitative agreement with our previous results for SRS approaching the integrability limit~\cite{Yash1}. Most important, we confirm that the RW appearance is independent of a particular source correlation model, thereby representing a universal signature of the self-similar regime of integrable turbulence. 
\begin{figure}[t]
\centering
\includegraphics[width=\columnwidth]{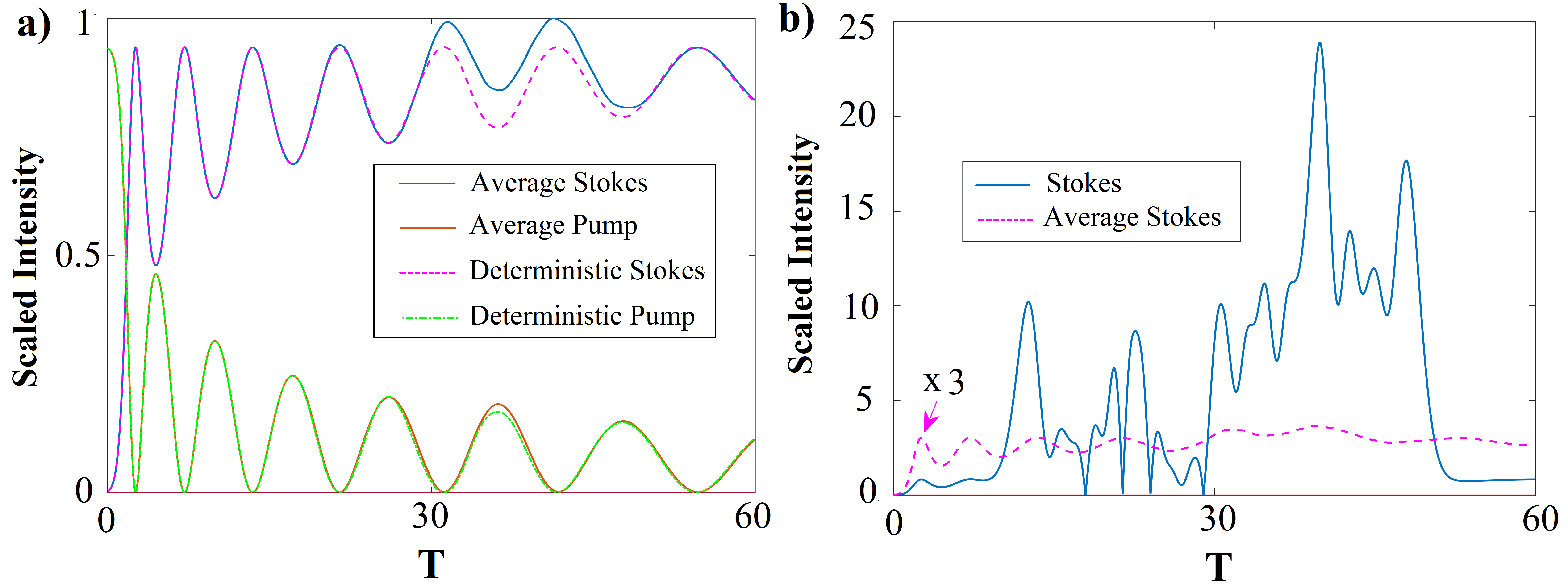}
\caption{(color online). (a) Scaled average intensities of the GSM Stokes (solid blue) and pump (solid red) pulse ensembles at $Z=3$. The scaled intensities of deterministic Stokes (dashed magenta) and pump (dash-dotted green) pulses 
in the absence of source noise are shown for comparison as well. (b) Scaled intensity of a random realization of the GSM Stokes pulse ensemble (solid blue) and the average Stokes pulse ensemble intensity (dashed magenta) at $Z=3$.  The average intensity is enhanced by a factor of three to facilitate visualization. All intensities are scaled to the total average intensity at the source. The numerical parameters are: $\kappa=1$, $\Delta P/P_0=0.4$, and $T_c=0.1T_{\ast}$.}
\label{figure4}
\end{figure}
\begin{figure}[t]
\centering
\includegraphics[width=\columnwidth]{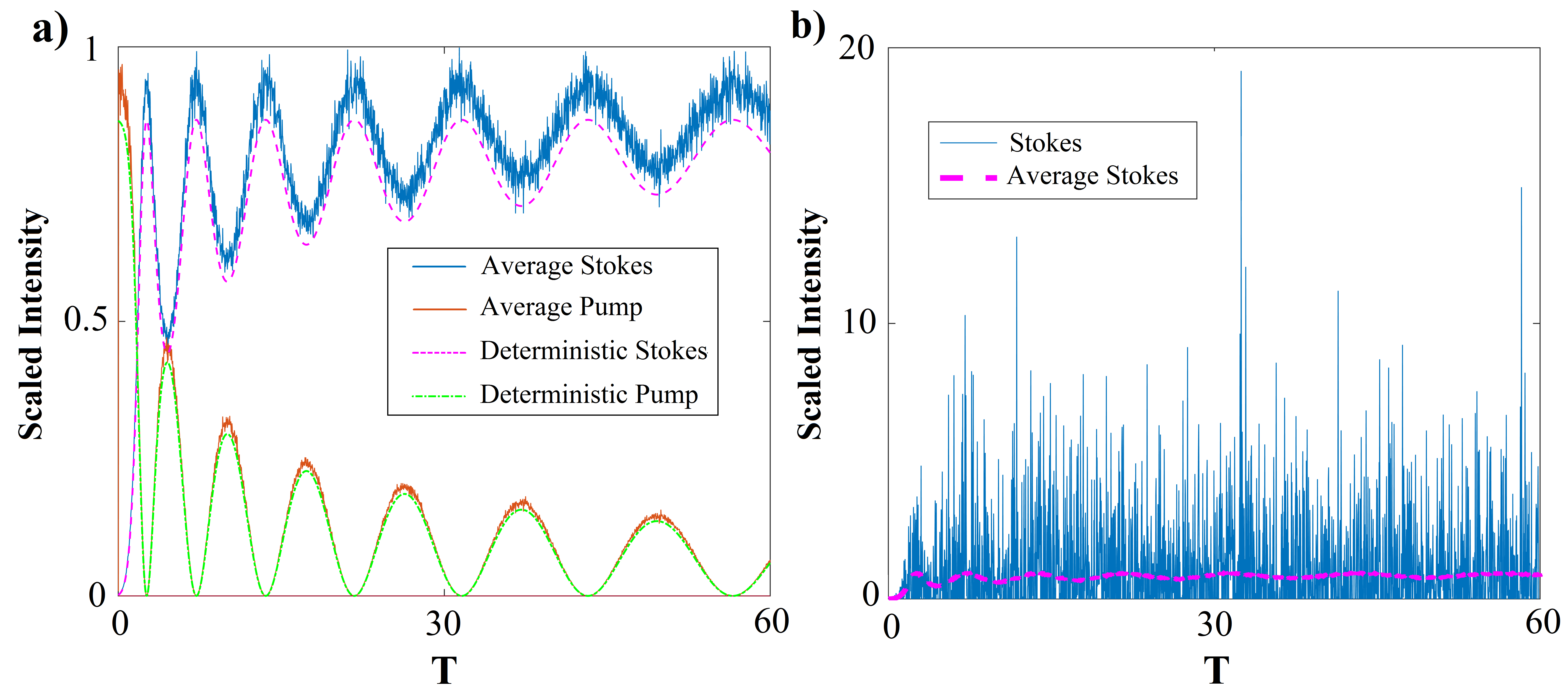}
\caption{(color online). Same as in Fig.4 for FBWN pulse ensembles.}
\label{figure5}
\end{figure}

To explore generic features--independent of a particular pump pulse profile at the source--of extreme event generation in the self-similar regime of integrable turbulence,  we scale away the source intensity profile by introducing the Stokes and pump pulse intensities scaled to the average total intensity at the source, $\lgl I_{\mathrm{tot}}\rgl$.  In Fig. 4a, we display the average dynamics of the scaled Stokes (solid blue curve) and pump (solid red curve) intensities for the GSM source ensemble, together with the corresponding deterministic SRS quantities for which no pump noise is present at the source. The deterministic dynamics of the scaled intensities of the Stokes and pump pulses are displayed with dashed magenta and dash-dotted green curves, respectively. We clearly observe that the average pulse dynamics follows closely the deterministic self-similar evolution scenario. Next, we juxtapose in Fig. 4b the self-similar average Stokes pulse intensity evolution in the scaled variables (dashed magenta curve) with a Stokes pulse ensemble realization dynamics in the scaled variables (solid blue curve). Note that the scaled average intensity is enhanced by a factor of 3 to facilitate visualization. It is evident from the figure that extreme events appear as giant fluctuations away from the average self-similar evolution of the system. Further, we repeat the calculations for the FBWN ensemble and exhibit the corresponding results for a white noise pump source in Fig. 5 using the same colour scheme. Despite quantitative differences of Fig. 5 from Fig. 4, our main conclusion regarding the RW manifestations as enormous fluctuations away from the long-term self-similar evolution of integrable turbulence holds for the FBWN source ensemble as well, supporting the universality of the proposed scenario.

\begin{figure}[t]
\centering
\includegraphics[width=7.5cm]{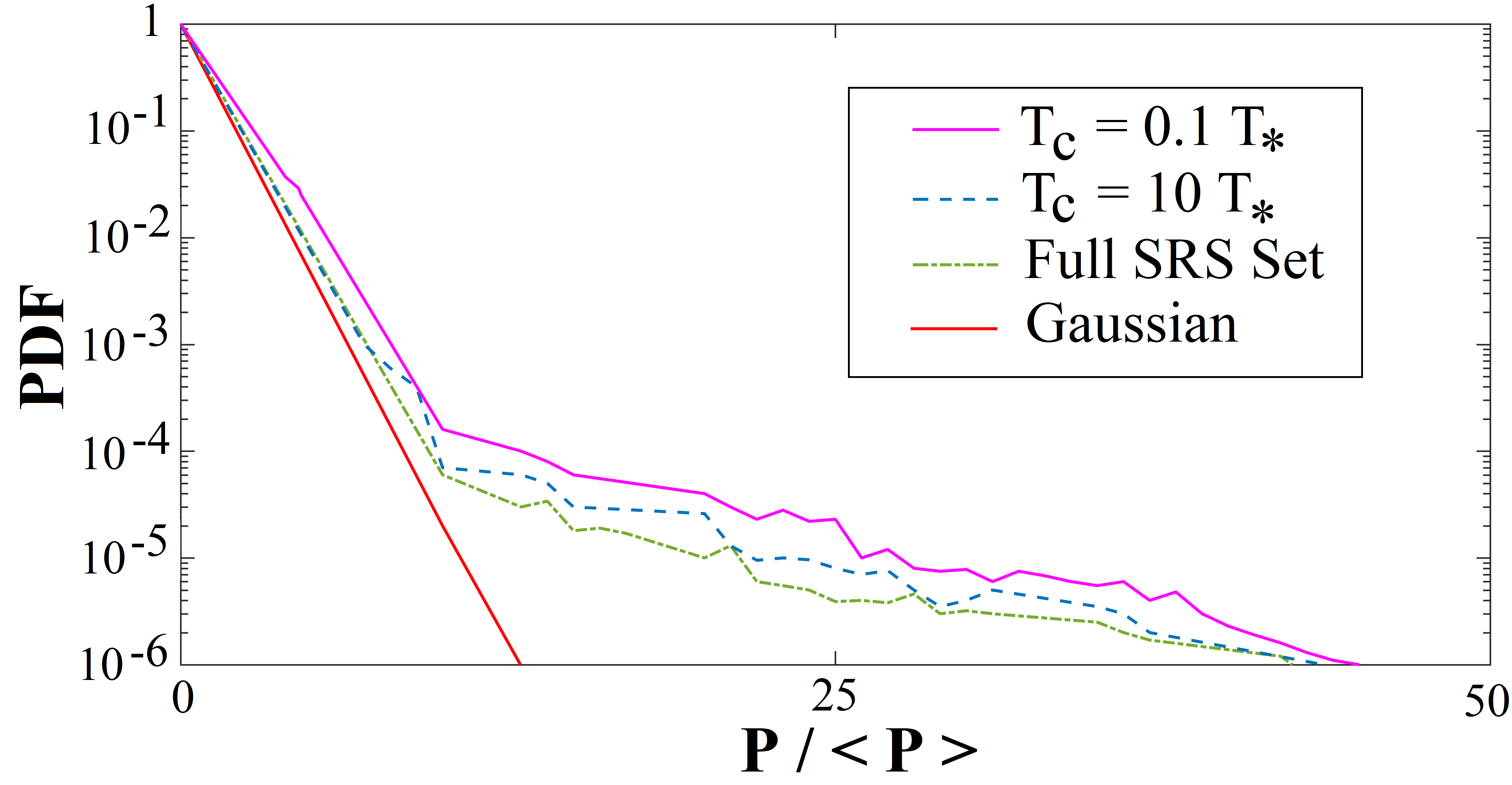}
\caption{(color online). Normalized peak power PDF of a GSM Stokes pulse ensemble at $Z=3$ for a nearly incoherent $T_c=0.1 T_{\ast}$  (solid magenta), and fairly coherent, $T_c=10 T_{\ast}$ (dashed blue) ensembles; the solid magenta and dashed blue curves are obtained using the self-similar evolution equations, Eqs.~(\ref{SRS-pst}) through~(\ref{SRS-SS}). The dash-dotted green curve displays the peak power PDF for a fairly coherent ensemble, $T_c=10 T_{\ast}$, evaluated using the full set of SRS equations, Eqs.~(\ref{Max}) and~(\ref{Sch}). The peak power is normalized to its average value at $Z=3$. The solid red line represents a Gaussian PDF with the same average peak power. The other numerical parameters are: $\kappa=1$ and $\Delta P/P_0=0.4$.}
\label{figure6}
\end{figure}
Finally, we examine  extreme event statistics by calculating the probability density function (PDF) of the normalized peak power of the Stokes pulse at an output distance. The peak power is normalized to its average value at the output distance. The results are shown in Fig. 6 where we display the peak power PDFs of the Stokes pulse output at $Z=3$ for a nearly incoherent GSM ensemble of pump pulses at the source, with the coherence time being the fraction of the input pulse width, $T_c=0.1 T_{\ast}$ (solid magenta curve), and that for a highly coherent GSM ensemble of such pulses with $T_c=10 T_{\ast}$ (dashed blue curve). The solid red straight line corresponds to a Gaussian PDF with the same average peak power as the average peak power of the Stokes ensemble at $Z=3$. We immediately notice that both output PDF curves strongly deviate from the Gaussian by acquiring long tails for the peak powers substantially exceeding their average values. This circumstance points to a greatly enhanced  likelihood of extreme event generation in the system, compared against predictions based on Gaussian statistics. We also observe that the solid magenta and dashed blue curves are very close to each other--taking into account inevitable data spread in a Monte-Carlo simulation--reinforcing our message that extreme event excitation in integrable SRS  is barely affected by the source coherence time. This conclusion is in qualitative agreement with our previous findings for extreme event excitation in non-integrable regime of SRS with noisy input pump~\cite{Yash1}.  The solid magenta and dashed blue curves were determined using the self-similar evolution description of SRS with the aid of Eqs.~(\ref{SRS-pst}) through~(\ref{SRS-SS}). To confirm that the system is indeed in the self-similar regime, we repeated the PDF calculations using the full set of SRS equations, Eqs.~(\ref{Max}) and~(\ref{Sch}). The resulting PDF is represented by a dash-dotted green curve for a highly coherent GSM ensemble with $T_c=10 T_{\ast}$ . We can clearly see in the figure that the dash-dotted green curve nearly coincides with the dashed blue curve, as expected, and is very close to the solid magenta one. It follows that the self-similar evolution description accurately captures the system statistics behaviour.

\noindent{\emph{Conclusion}.--}We have demonstrated that RWs can be excited in a self-similar asymptotic regime of integrable turbulence and they appear as giant fluctuations away from the average (self-similar) evolution of the system. Although our results are based on numerical simulations of SRS in the transient regime, the conclusions are model independent; we can expect qualitatively similar findings in the TLA case.  Indeed, the TLA and SRS in the integrable regime can be shown to be governed by the same set of evolution equations with an appropriate choice of variables~\cite{Chu}.  However, there is an important difference: While one can neglect the amplified spontaneous emission noise contribution to SRS in gases in the weak excitation approximation--because there is a negligible population of the excited virtual states of the two-photon transition--strongly amplified spontaneous emission noise cannot be neglected in TLA. We anticipate that the additional amplified noise in the TLA case will only increase the likelihood of extreme event occurrence in the system. We plan to examine this point in detail elsewhere.

\end{document}